# Compression failure of porous ceramics: A computational study about the effect of volume fraction on the failure behaviour

Vinit Vijay Deshpande[1] and Romana Piat[1]

[1] Department of Mathematics and Natural Sciences, University of Applied Sciences Darmstadt, Schöfferstraße 3, Darmstadt 64295, Germany

**Abstract:** The work describes a numerical method to study the nature of compressive failure of porous ceramics in relation to its volume fraction. The microstructure of an alumina-based foam material manufactured by mechanical stirring of a slurry is studied here. A finite element based compressive failure simulation of a real microstructure obtained from microtomography (µCT) scans is conducted. A recently developed microstructure reconstruction algorithm is utilized to generate artificial microstructures statistically equivalent to the real one obtained from µCT. The accuracy of the reconstruction procedure is established by comparing the simulated compression behaviour of the reconstructed microstructure with that of the real one along with the experimentally measured results. The effect of sample size on the simulated compression behaviour is studied by computing compression stress-strain behaviour for varying sizes of the reconstructed microstructures. Further, artificial microstructures of the porous ceramic with different volume fractions are reconstructed along with computing compression stress-strain behaviour to establish relationship between ceramic content (volume fraction) and compressive strength of this material. The nature of the compressive failure for microstructures with different volume fraction is studied and the results are compared with the analytical models and the experimental observations available in the literature.

## 1. Introduction

Porous ceramics form a very special category of materials as they inherit the properties of their base ceramic material while being extremely light weight due to the porosity. Ceramics have high melting point, high strength and high corrosive and wear resistance. Making them porous imparts additional properties like low density, low thermal conductivity, high surface area and high specific strength [1]. Due to the brittle nature of the ceramics, inserting pores in ceramics had been considered a bad idea historically. However, in the last few decades, wide applications of porous ceramics have emerged especially for high temperature and corrosive applications. Some examples of its applications include high temperature insulation, filtration of molten metals, particulate matter filtration from exhaust gases, light weight construction materials, bone substitutes etc.

Compression strength is an important property for porous ceramics as they are subjected to significant compressive stresses during applications like metal filtration, construction, bone substitutes, etc [2]. Therefore, studying the effect of porosity (or ceramic volume fraction) on the compression strength and the compressive failure behaviour in general is pertinent for their improved usage in such applications. Limited experimental data exists in literature that studies the failure mechanisms involved in compression failure of ceramic foams and the effect of volume fraction on the compressive behavior. [3, 4] performed compression strength measurements on silicon oxycarbide foams having volume fraction in the range of 0.15 – 0.3 produced by direct foaming of preceramic polymer. The foams displayed high dimensional stability and consistent compression strength at elevated temperatures. [5] measured the compression strength of two alumina foams having volume fraction 0.05 and 0.1. [6] studied the effect of loaded area, rate of applied load and pore size of alumina, ZrO$_2$, SiC and fused silica foams on their compression strengths. They observed that the rate of applied load has no significant influence on compression strength of ceramic foams. However, the loaded area and sample size have noticeable influence. [2] experimentally observed the compression failure behaviour of alumina foams with volume fraction in the range of 0.25 to 0.7. Through tomographic and SEM images, it concluded that foams below 0.5 volume fraction displayed cellular failure (progressive damage) while those above that displayed brittle (abrupt) failure. Graphite foams were studied in [7] which described the effect of boundary conditions on the compression failure behaviour. It also described the dependence of compression strength on sample size of the test specimens.

Theories describing the compression behaviour of porous materials can be distinguished into two major works. The formulation in [8], [9] dealt with behaviour of isolated pore in a finite sized specimen subjected to compressive stress. This was then further extended to multiple pores distributed inside the specimen. The relationships between stress intensity factors and crack lengths for these configurations were devised followed by creation of failure surfaces in stress space. This theory is applicable to materials with low porosity content or a smaller number of pores. [10], [11] developed proportionality laws that described relationships of effective material properties like stiffness, elastic buckling, plastic collapse stress and crushing strength with the relative density (volume fraction) for ductile and brittle foams. The laws were based upon the idea that the effective material properties of the foam are related to the mechanics of bending, buckling, plastic collapse and brittle fracture of the cell walls (in case of closed-cell foams) and cell edges (in case of open-cell foams). This theory is applicable to materials with a very high content of porosity. By studying the 2-dimensional honeycomb structure, expressions describing the effect of sample size on compression strength of cellular materials were developed in [12] that concluded that compression strength increases with increase in sample size till a representative size is reached. This was validated through experimental measurements on aluminium foams in [13]. This study was extended in [14] which concluded that the size dependence of compression strength arises from the inherent structural defects in the material which otherwise would not exists beyond a certain sample size.

The reference material chosen in the present work is an alumina foam reported in [15]. It was manufactured by mechanical stirring of alumina slurry and had highly homogeneous microstructure with almost spherical shaped pores. The detailed description of the manufacturing and the measurement of the effective elastic properties can be found in [15]. The stress - strain behaviour of the foam when subjected to compressive loading was studied in [16] which described the nature of crack propagation and the mode of sample failure.

Our recent article [17] studied the foam material described above with an objective of performing numerical microstructure characterization and determination of effective elastic properties. The random nature of the foam microstructure required a large number of foam samples for performing any useful simulations. To achieve this, a numerical microstructure reconstruction algorithm was developed that can generate statistically equivalent microstructures for a given distribution of pore sizes and volume fraction. Using this procedure, artificial microstructures were generated that were equivalent to a real one in terms of their statistical correlation functions. Through finite element simulations, effective elastic properties of the material were calculated by homogenizing the reconstructed and the real microstructures. The properties obtained from both the microstructures were matching to each other and fared well in comparison with the experimental results [15] as well.

The objective of the present work is to study the compression failure behaviour of porous ceramics with respect to their volume fraction. To do this, firstly finite element simulations are performed on the microstructure of real foam material studied in [15], [17] to study its compressive failure behaviour, crack propagation and sample failure mechanism. Then the reconstruction procedure developed in [17] is used to generate artificial microstructures and their compressive failure behaviour is simulated and compared with that of the real microstructure. Having proved the fidelity of the reconstruction procedure, artificial microstructures with different volume fractions ranging from 0.125 to 0.875 are developed and the simulation procedure repeated to study the effect of volume fraction on the compression failure behaviour. Note that for each volume fraction, ten artificial microstructures each of three different sample sizes are created and simulated to study the scatter in compression stress-strain results with respect to the sample size. The simulation results are then compared with Gibson – Ashby model [10] as well as experimental observations available in the literature.

The article is organised into following sections. Section 1 introduces the topic of research, describes the state of the art and the summary of the procedure adopted in this paper. Section 2 describes a real foam microstructure, details the simulation procedure of the compression failure, briefly mentions the microstructure reconstruction procedure and finally describes the scatter in the compression failure behaviour with respect to sample size. Section 3 describes reconstruction algorithm for microstructures with different volume fractions, scatter in the compression failure behaviour, energy absorption behaviour and nature of damage propagation with respect to volume fraction. Section 4 discusses all the results and draws inferences while section 5 concludes the topic.

## 2. Study of real alumina foam

The alumina foam manufactured in [15] is chosen as a reference material system in this work so as to generate high-fidelity numerical simulations. Fig.1a shows a 3D binary image of the foam microstructure obtained by processing the micro-tomographic images. The black color region represents pores whereas the white color is alumina. The volume fraction of alumina content obtained from experimental measurements [15] was 25.5 %. It was observed that the shape of pores was almost spherical with size distribution as shown in Fig. 1b. Since the sample size shown in Fig.1a was too big for finite element simulations, an appropriate size and position of statistical volume elements from within the entire sample were determined in [17] and further validated by comparing the linear elastic properties obtained from finite element simulations with that of the experimental results [15].

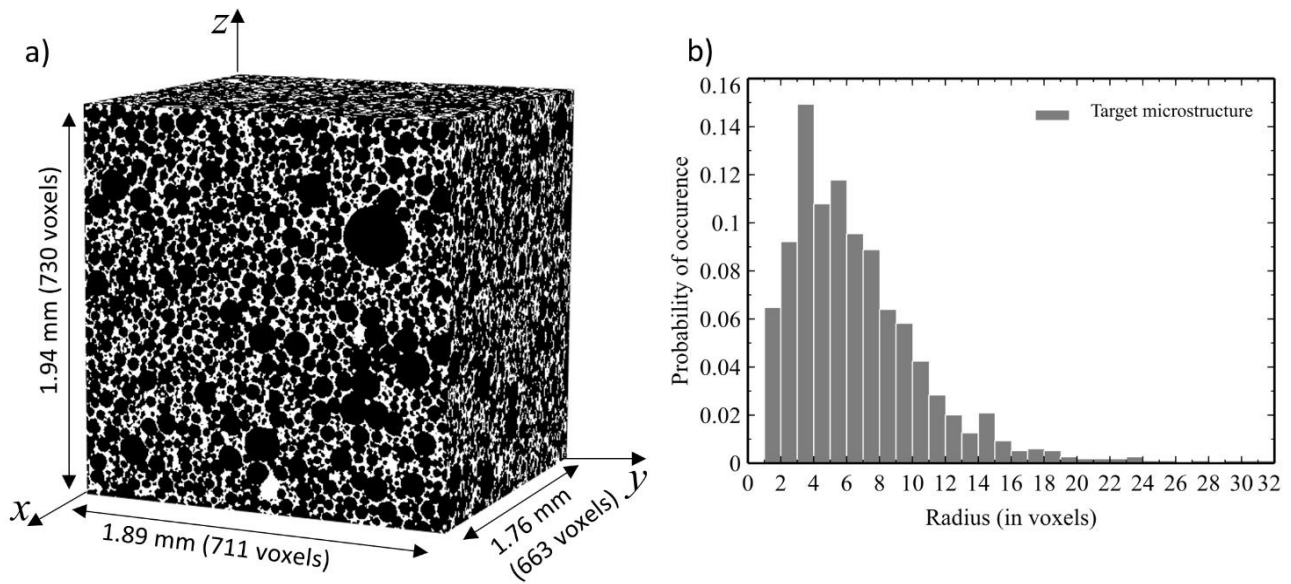

Fig. 1: a) Microstructure of alumina foam; b) size distribution of pores in the microstructure.

## 2.1. Simulation of compression failure of a real foam sample

In order to study compression failure behaviour of the foam sample, a volume element (VE) obtained from studies in [17] is chosen as shown in Fig.2. The VE is in the form of a 3D binary image. A surface mesh with triangular elements is generated from the VE in MATLAB [18] followed by creation of a volume mesh in Abaqus [19]. Linear tetrahedral elements are chosen for the purpose of mesh creation. The constitutive behaviour of the alumina base material is modelled using Johnson Holmquist-2 (JH-2) material model [20] which is particularly suited for modelling damage in brittle materials. All the material parameters for the JH-2 model of alumina are adopted from [21] except for maximum fracture strength and shear modulus (refer Table.1) which are tuned so that the compression strength obtained finite element simulations correspond to that obtained experimentally in [22]. Fig.2 shows the microstructure volume element of the foam sample subjected to boundary conditions. The bottom face of the VE is fully constrained while the top face is allowed to move only in the normal direction. A compressive strain ($\varepsilon_z = -0.7\%/s$ in Fig.2) is applied to the top face in a quasi-static analysis in Abaqus. The damage in the material is modelled by deleting the finite elements in accordance with the progressive damage criteria of the JH-2 model.

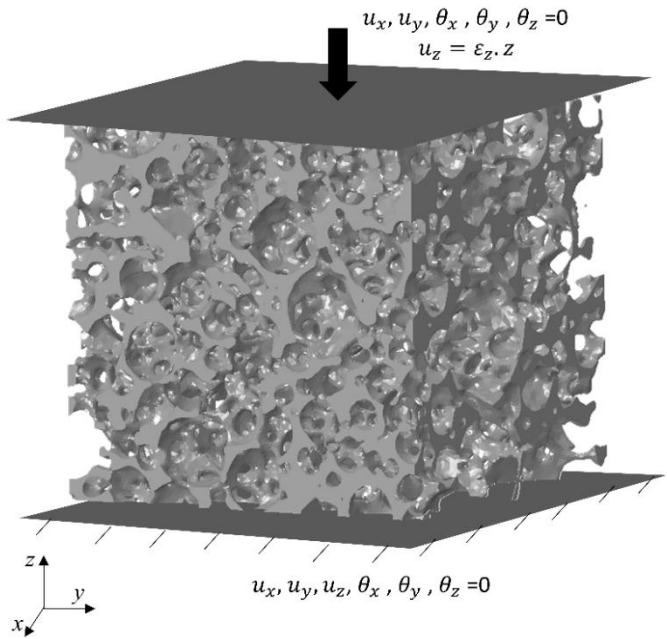

| Alumina material parameters | Values |
|---|---|
| Shear modulus (GPa) | 190 |
| Normalized maximum fracture strength | 0.92 |

Table 1: Alumina material parameters obtained from tuning process.

Fig. 2: Microstructure volume element and boundary conditions.

Fig.3 shows effective compression stress-strain curve obtained from the displacement and the reaction force calculated at the top side of the VE. The simulation is performed along three orthogonal directions as indicated in the figure. The three curves without makers are the experimentally obtained curves on three different specimens mentioned in [22]. The experimental data was available only till the point at which the maximum stress was reached. The average maximum stress from the three simulation curves is 65.2 MPa while that from the three experimental curves is 65.7 MPa. According to the simulation data, the maximum compressive stress is reached at the strain value of 0.33%.

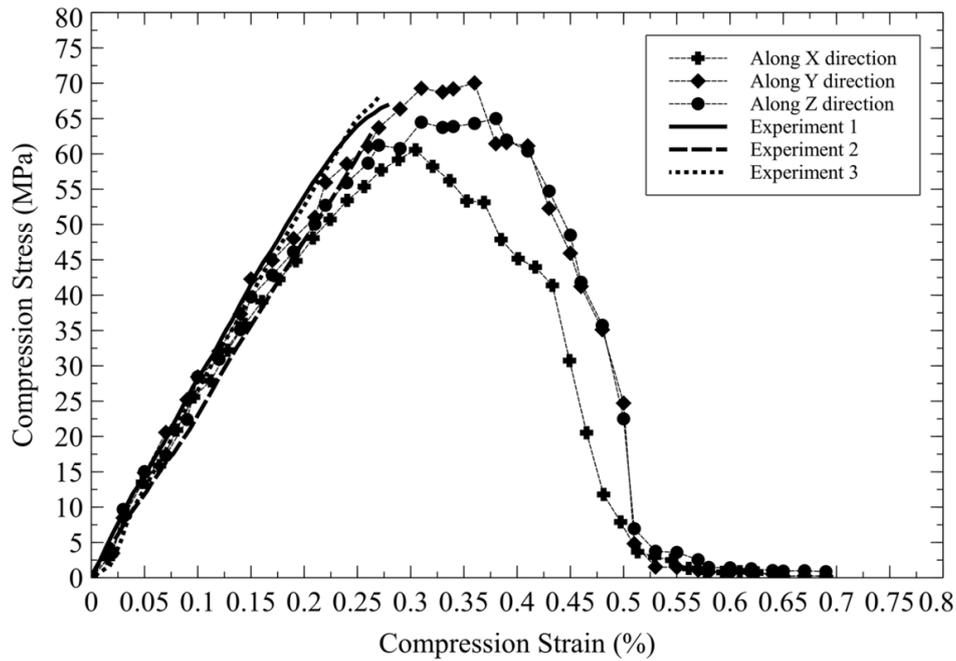

Fig. 3: Effective compression stress-strain curves from simulation along three orthogonal directions and the three experimentally measured curves [22].

Fig. 4a shows compression stress-strain curve along X-direction (for coordinate system please refer Fig. 2). The direction of loading and the final damage pattern at the end of simulation in shown in Fig. 4b. In order to study the nature of damage propagation, five intermediate stages of the simulation course, marked as A to E are shown in Fig. 4a. As the compressive strain is increased, the microstructure undergoes elastic deformation. The formation of first damaged region is observed at stage A at the strain value of 0.08%. Its location can be seen in Fig. 4c in red color on one of the struts (cell walls) in between the pores. As the strain is increased, damage gets initiated at multiple struts in the microstructure. Fig. 4d shows the damage pattern (red color regions) at stage B at the strain value of 0.2%. From the damage pattern distribution, it can be seen that the damage initiation is not localized but distributed throughout the material region. The maximum compressive stress is reached at stage C at the strain value of 0.3%. The damage pattern at this stage can be seen in Fig. 4e. At this stage, damage has been initiated at multiple new struts. The pre-existing red regions have grown in size indicating that some of the old struts where damage had initiated before have been severely damaged. The damage pattern at stage D at the strain value of 0.43% can be seen in Fig. 4f. At this stage, a significant number of struts have been completely damaged which has resulted in noticeable loss of sample strength. Stage E at the strain value of 0.65% shows total loss of sample strength. Fig.4g shows the damage pattern at this stage which indicates completely damage struts. The same damage pattern along with undamaged foam region can be seen in Fig. 4b. The sample at this stage is broken into multiple smaller pieces.

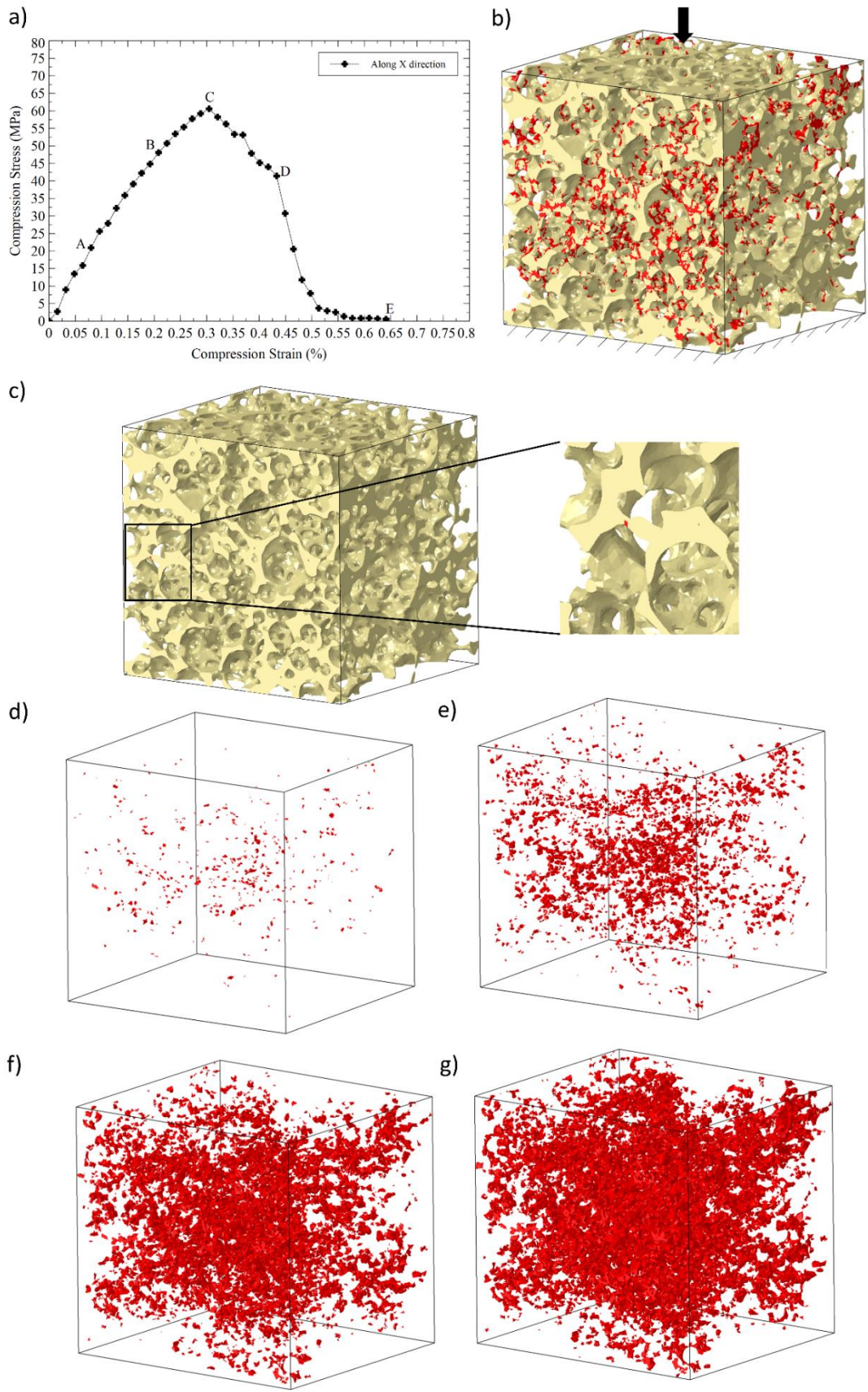

Fig. 4: a) Effective compression stress-strain curve of VE of a real preform sample; b) final damage pattern in VE; c) location of initial damage in the VE; damage pattern at d) stage B, e) stage C, f) stage D and g) stage E of the simulation course.

## 2.2. Reconstruction of artificial foam samples

In our previous article [17], an important objective was to perform statistical analysis on the microstructure parameters and the effective elastic properties of the alumna foam. To achieve this, a microstructure reconstruction algorithm was developed that could generate artificial microstructures statistically equivalent to the real one. It is an optimization algorithm based on Yeong-Torquato (YT) method [23] that utilized the peculiar microstructure of the foam to drastically reduce the number of iterations which would have otherwise required. Three correlation functions namely 2-point correlation function, 2-point cluster correlation function and lineal path function were calculated for the foam microstructure shown in Fig.1. In a multiphase system, 2-point correlation function calculates the probability of two points separated by a certain distance to lie in the same phase of interest. 2-point cluster correlation function calculates the probability of two points separated by a certain distance to lie in the same cluster (connected region) of the phase of interest. Lineal path function calculates the probability of an entire line segment of a certain length to lie in the phase of interest. The graphs of these functions for the chosen microstructure are shown in Figs.5a-c. The detailed description of these functions along with their calculation methods can be found in [17].

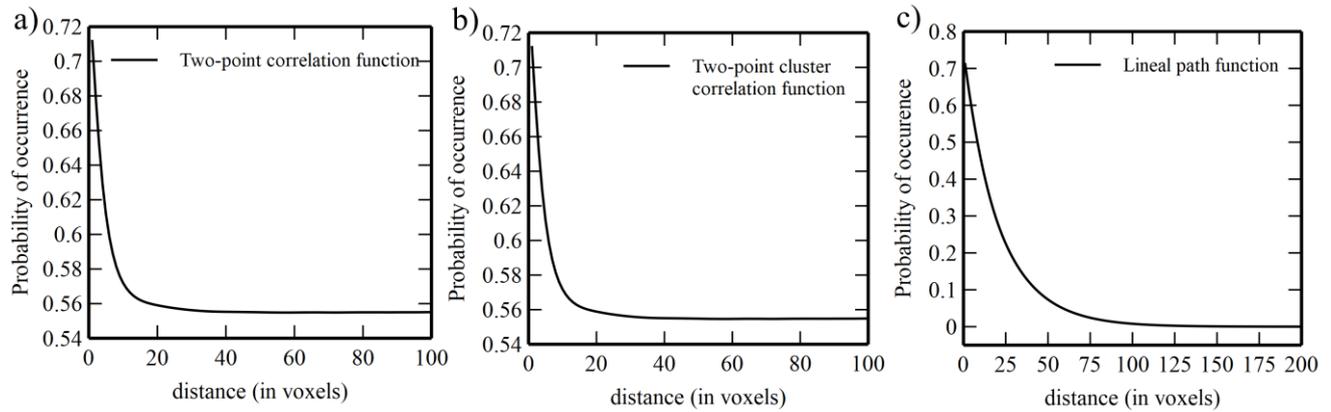

Fig. 5: a) 2-point correlation function, b) 2-point cluster correlation function and c) lineal path function of the alumina foam microstructure.

The objective function of the optimization algorithm is defined as energy functional, $E$ in Eq. (1).

$$E = \sum_\alpha w_\alpha \sum_r [f(r) - \bar{f}(r)]^2 \tag{1}$$

$f(r)$ and $\bar{f}(r)$ indicate statistical correlation functions dependent on distance $r$ in iterated and target microstructures respectively. $\alpha$ indicates number of correlation functions defined in the energy functional ($\alpha = 3$ in this case) and $w_\alpha$ indicates weight assigned to each function. The real foam microstructure in Fig. 1a was chosen as a target microstructure. The iterations in the optimization algorithm proceed till the statistical correlation functions of the iterated microstructure match to that of the target microstructure. The definition of initial microstructure and the method of perturbation differentiates this algorithm from the YT method. Instead of choosing an initial microstructure as a random distribution of voxels as described in the YT method, this work utilizes the available information of shape and size distribution of pores to define a much more realistic initial microstructure. Further, the perturbation in each iteration is carried out not by swapping black and white voxel positions (as in YT method) but by changing the location of spherical pores. This results in the method requiring much smaller number of iterations. Detailed description of the method can be found in [17]. Figs. 6a-b show binary images of the real and reconstructed microstructures. The reconstruction algorithm gives us the ability to generate microstructure volume elements of different sizes that are statistically equivalent to the real microstructure. Through this we can study the scatter in the compressive failure behaviour of the material with respect to the VE size.

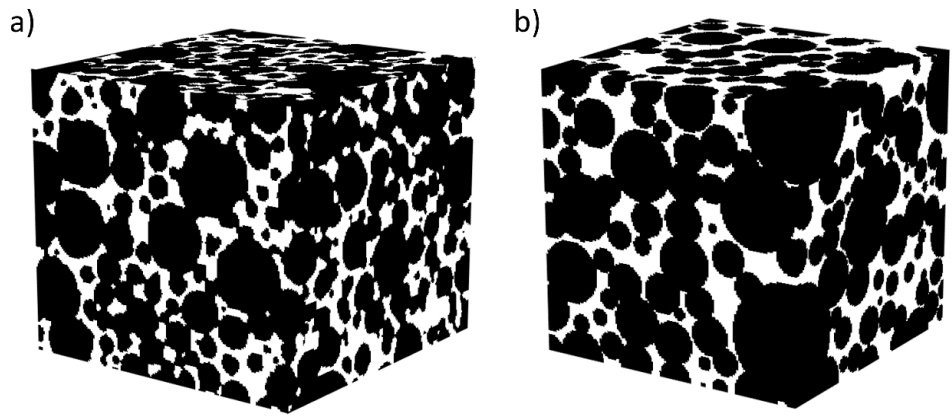

Fig. 6: Binary 3D image of a) real and b) reconstructed microstructure of the alumina foam.

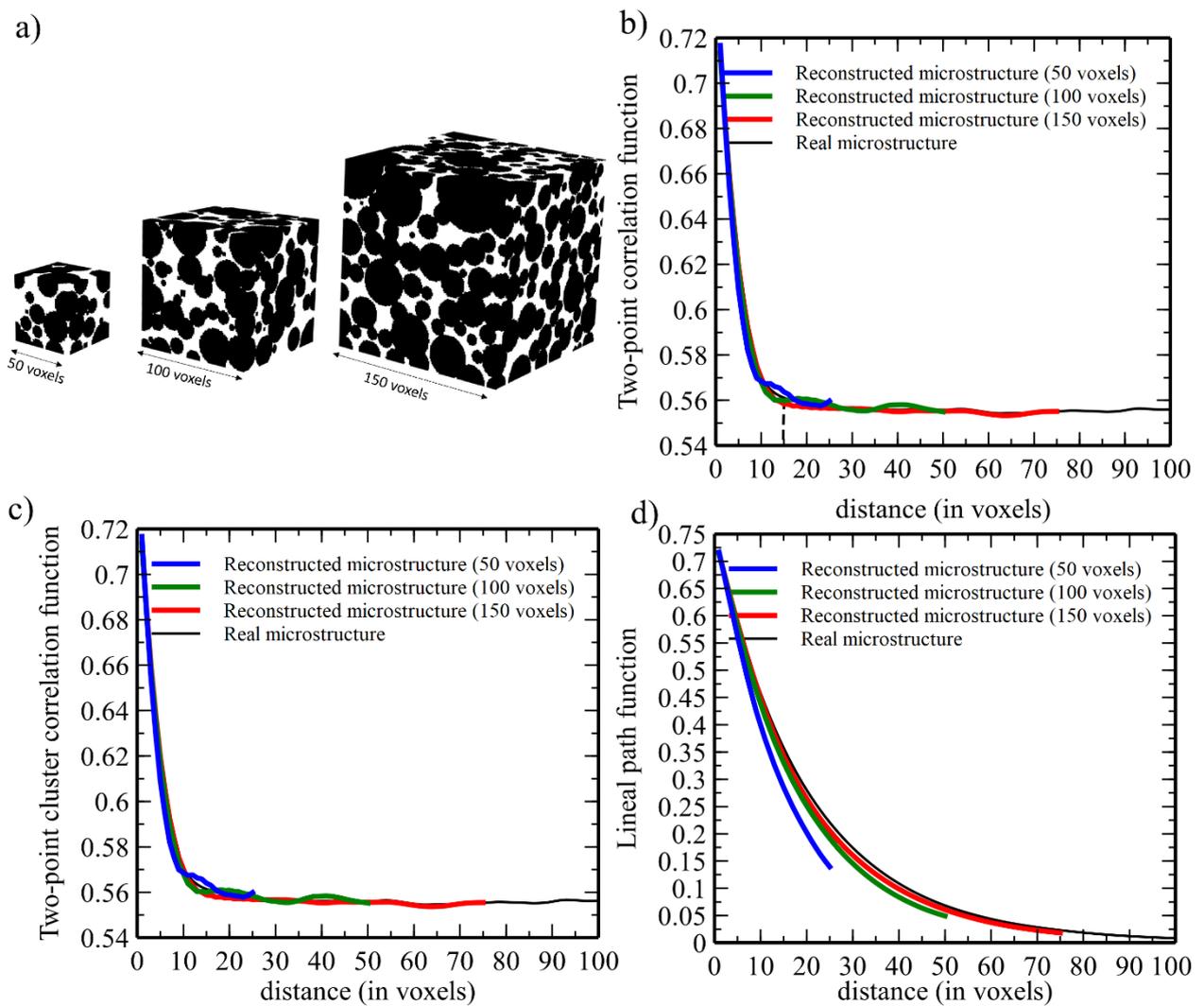

Fig. 7: a) Reconstructed volume elements with edge length 50, 100 and 150 voxels; b) 2-point correlation function, c) 2-point cluster correlation function and d) lineal path function of volume elements with three different edge lengths and the target microstructure.

To do this, three different sizes of the VEs with edge length 50, 100 and 150 voxels are generated through the reconstruction algorithm. The VEs with three different sizes are shown in Fig. 7a. Figs.7b-d show the three statistical correlation functions of the reconstructed microstructure of each size along with that of the real foam microstructure. Note that since these functions indicate probabilities measured in a finitely spaced region, the accuracy of the calculation depends on the distance $r$ while measuring the function $f(r)$. The accuracy reduces at distances close to the dimensions of the studied region. Hence, while calculating correlation functions, the maximum value of $r$ is considered to be half of the edge length of each VE.

**2.3. Simulation of compression failure of artificial foam samples**

For each selected size, ten microstructure VEs are generated through the microstructure reconstruction algorithm. Each of these VEs are converted into finite element models according to the procedure described for the case of real foam sample in section 2.1. Each VE is simulated for the compression failure behaviour along each of the three orthogonal directions. In order to study the scatter in the compression stress-strain behaviour, shadow curves are plotted for each VS size in Fig.8a. Each shadow curve is formed by enveloping the three stress-strain curves (for three directions) of all the ten VEs of each VE size.  So, each shadow curve envelops thirty stress-strain curves. The resulting shadow curves along with the stress-strain curves of the real foam sample are plotted in Fig. 8a. It can be seen that the scatter reduces as the size of VEs is increased with the scatter converging towards the simulation curves of the real foam sample.

Next, the characteristic length of the foam microstructure is calculated as the length at which the 2-point correlation function becomes long-ranged. This value is identified as 15 voxels (refer the dotted line in Fig. 7b). Further, a dimensionless term 'size factor' is defined as shown in Eq. (2) to represent the length scale of the microstructure VEs. The compression strength obtained from each of the thirty stress-strain curves along with their average value is plotted in Fig. 8b as a function of the size factor.

$$Size\ factor = \frac{edge\ length\ of\ volume\ element}{characteristic\ length\ of\ microstructure} \quad (2)$$

Fig. 8b shows that the average compression strength for the VE size of 150 voxels (size factor =10) is 70.7 GPa. Fig, 9a-b shows final damage pattern in the real (Fig.9a) and the largest reconstructed (Fig. 9b) microstructure.

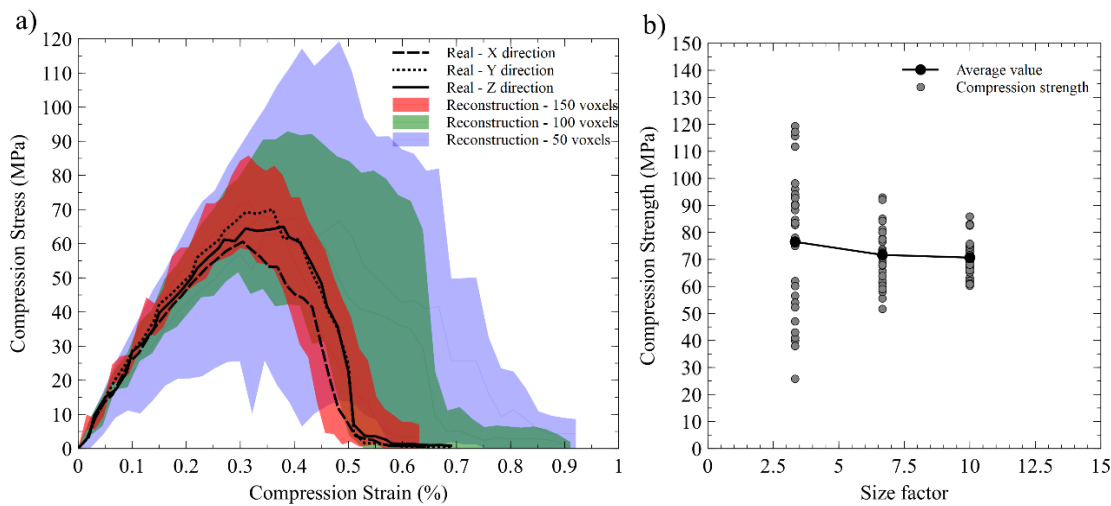

Fig. 8: a) Shadow curves of compression stress-strain behaviour for 10 reconstructed VEs for each edge length of 50,100 and 150 voxels along with stress-strain curves obtained from simulation of real foam sample.

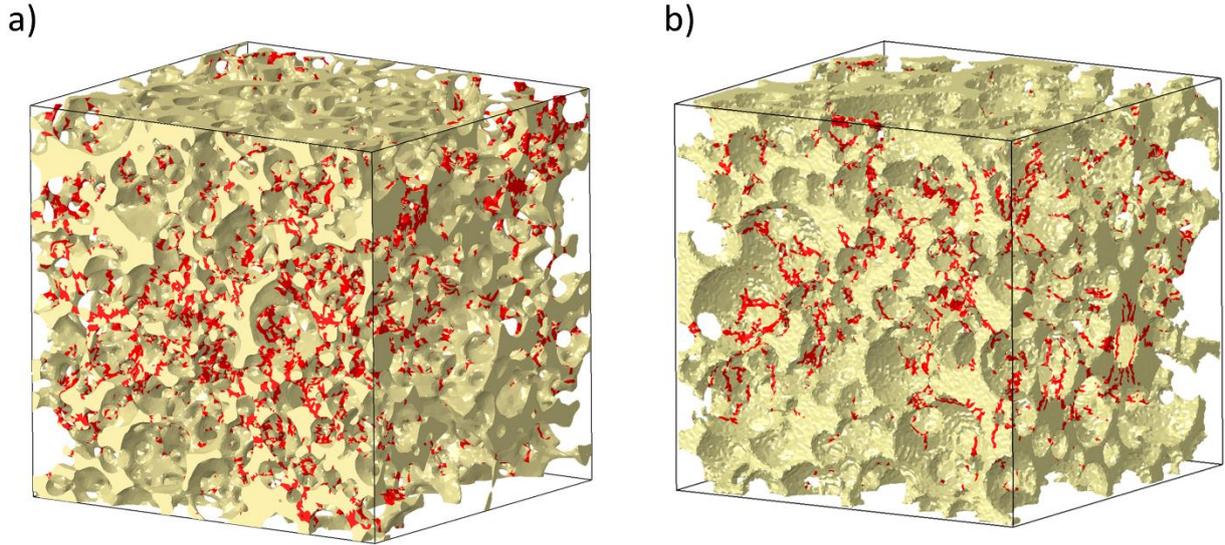

Fig. 9: Final damage pattern in a) real and b) reconstructed (edge length 150 voxels) microstructures of alumina foam.

## 3. Study of alumina foam with different volume fractions

In order to study foams with different volume fractions, the first step is to generate artificial microstructures. The microstructure reconstruction algorithm requires volume fraction, shape and size distribution of pores, size of volume elements and target statistical correlation functions as inputs. In this study, we have assumed that the shape and size distribution of pores remains the same (refer Fig. 1b) irrespective of the volume fraction. For each volume fraction, three sizes of volume elements are studied as described the section 2.2. The only remaining input is the objective function defined through target correlation functions.

### 3.1 Target statistical correlation functions

Since a 2-point correlation function $S_2(r)$ defines the probability of 2 points separated by a distance $(r)$ to lie in the same phase of interest, in the limit when $r \to 0$,

$$\lim_{r \to 0} S_2(r) = v_f \tag{3}$$

where, $v_f$ is the volume fraction of the phase of interest (pores in this case). Further, in case of a homogeneous microstructure without any long-range order,

$$\lim_{r \to \infty} S_2(r) = v_f^2 \tag{4}$$

In the case of the reference real foam microstructure with $v_f = 0.745$ (porosity), it can be seen in Fig. 10a, that both these definitions are satisfied. Hence, for a homogeneous foam microstructure with any other volume fraction, these two quantities of the $S_2(r)$ are readily known. The only unknown region of the function is the one highlighted in red color in Fig. 10a.

In order to determine this unknown region, a study is conducted on the real foam microstructure whose correlation functions are already known (refer Fig.1). An initial microstructure is generated for the reconstruction algorithm following the procedure described in [17] for volume elements of edge length 50, 100, 150, 200 and 250 voxels. $S_2(r)$ is calculated for each of these volume elements and compared with that of the real microstructure to calculate the error as defined in Eq. (5).

$$Error_{S_2(r)} = \sqrt{\frac{\sum_r(S_2(r)-\bar{S}_2(r))^2}{\sum_r r}} \tag{5}$$

Here, $Error_{S_2(r)}$ is the error associated with every $S_2(r)$ and $\bar{S}_2(r)$ is the correlation function of the real microstructure. It is observed in Fig. 10b that the error is reduced as the size of volume element is increased. Hence, the unknown region in Fig. 10a can be reasonably defined from the $S_2(r)$ of the initial microstructure of the volume element with edge length 250 voxels.

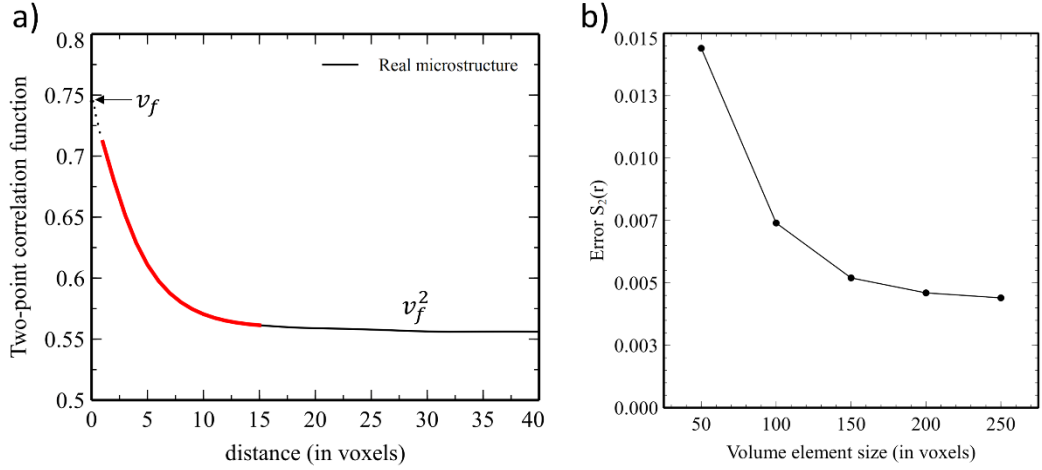

Fig. 10: a) 2-point correlation function of real foam microstructure; b) $Error_{S_2(r)}$ calculated from Eq. (5).

This procedure is followed to create target 2-point correlation function for microstructures with volume fraction 0.125, 0.375, 0.5, 0.625 and 0.875. 2-point cluster correlation function and lineal path function for these volume fractions are directly adopted from that of the initial microstructures with edge length 250 voxels. These functions for the selected volume fractions are shown in Figs. 11a-c.

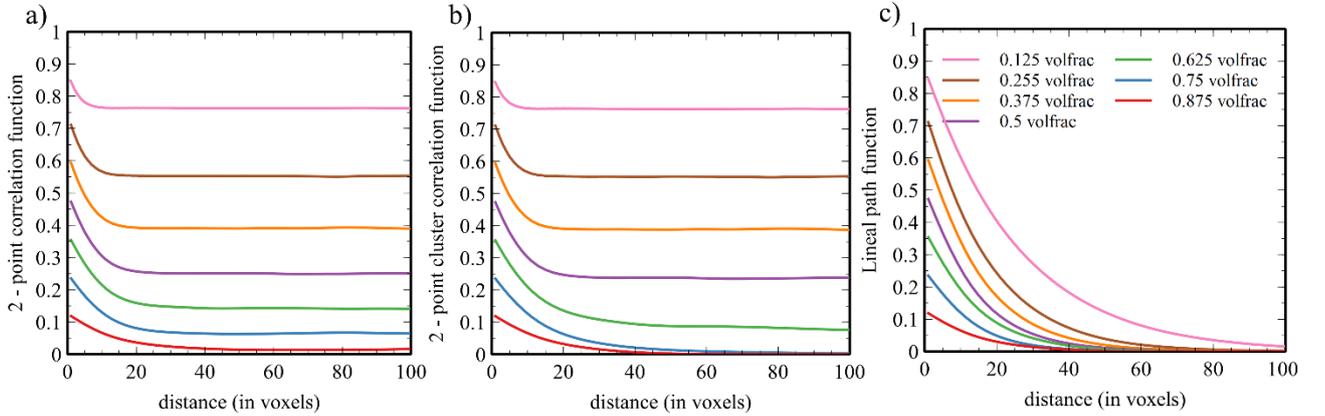

Fig. 11: Target a) 2-point correlation function, b) 2-point cluster correlation function and c) Lineal path function for all selected volume fractions.

### 3.2 Microstructure reconstruction

Now that all the inputs required for the microstructure reconstruction algorithm are available, artificial microstructures are generated for the selected volume fractions. Ten VEs each for edge length of 50, 100 and 150 voxels are generated for every volume fraction. 3D binary images of VEs with the largest edge length are shown in Fig. 12a-f. Further, characteristic length and size factor of VE of each size are calculated for every volume fraction according to the process described in section 2.3 and Eq. (2). Table. 2 shows characteristic length and size factor for VEs of all volume fractions.

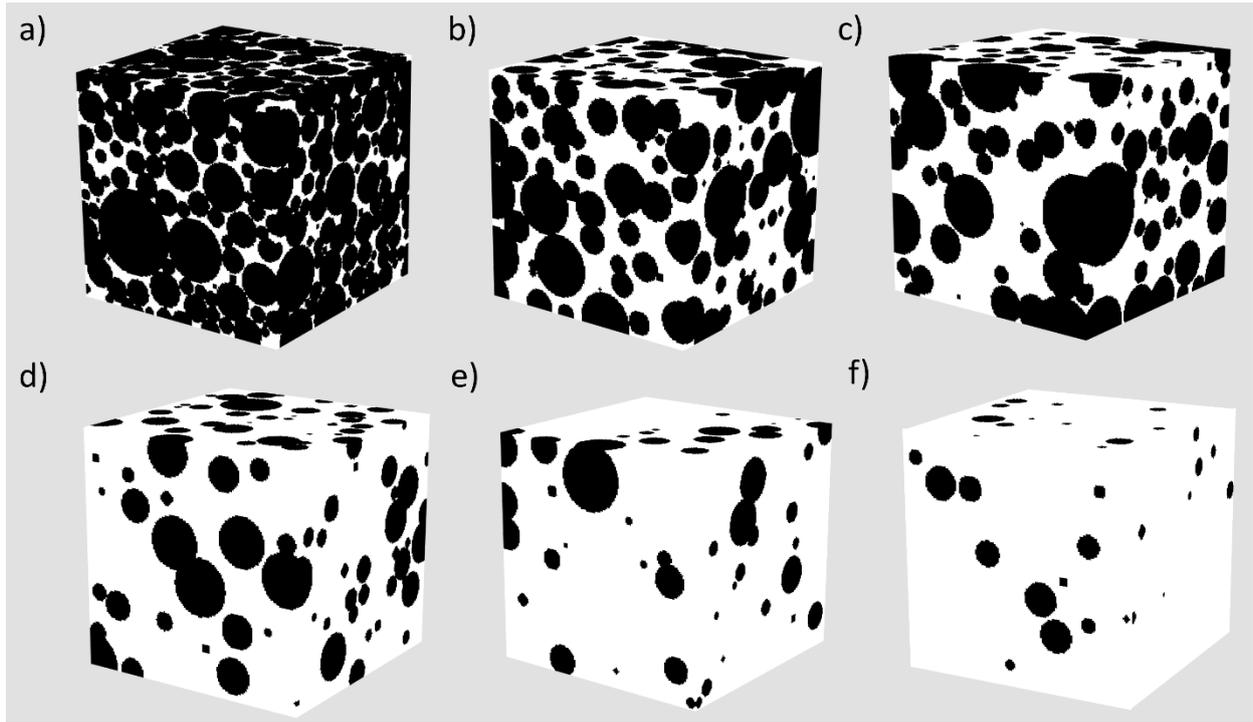

Fig. 12: Reconstructed VEs with edge length 150 voxels for volume fraction of a) 0.125, b) 0.375, c) 0.5, d) 0.625, e) 0.75 and f) 0.875.

| Alumina volume fraction of VE | Characteristic length (voxels) | Size factor | | |
|---|---|---|---|---|
| | | Edge length 50 (voxels) | Edge length (100 voxels) | Edge length (150 voxels) |
| 0.125 | 10 | 5 | 10 | 15 |
| 0.255 | 15 | 3.34 | 6.67 | 10 |
| 0.375 | 17 | 2.95 | 5.89 | 8.82 |
| 0.5 | 20 | 2.5 | 5.0 | 7.5 |
| 0.625 | 25 | 2 | 4 | 6 |
| 0.75 | 30 | 1.67 | 3.34 | 5 |
| 0.875 | 50 | 1 | 2 | 3 |

Table 2: Characteristic length and size factor for each edge length of VEs for every volume fraction.

**3.3 Simulation of compression failure of microstructures with different volume fractions**

The binary image of each VE is converted into a tetrahedral volume mesh according to the process described in section 2.1. Finite element simulation of the compression failure of each VE is performed followed by calculation of the shadow curves to study the scatter in the stress-strain behaviour with the size of VEs (refer left side images of Figs. 13a-f). Further, the scatter in the values of compression strength is studied with respect to the size factor (refer right side images of Figs. 13a-f).

a)
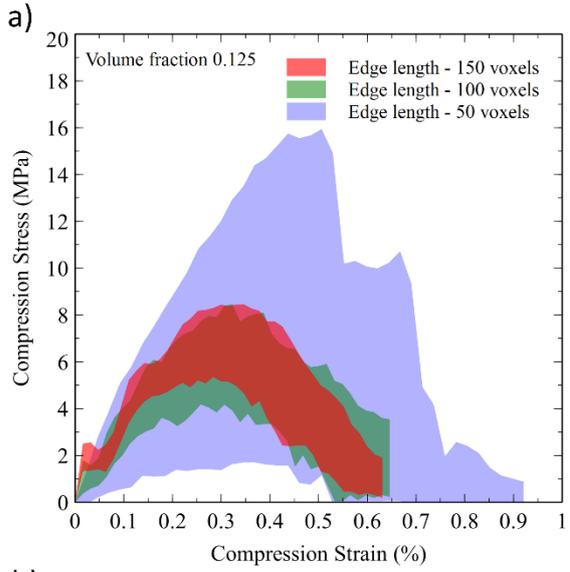
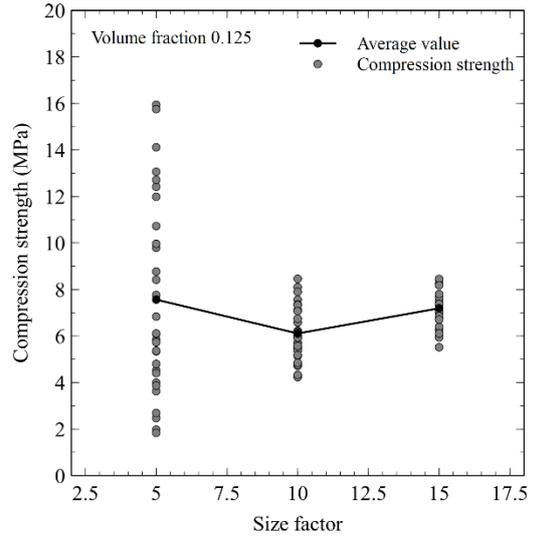

b)
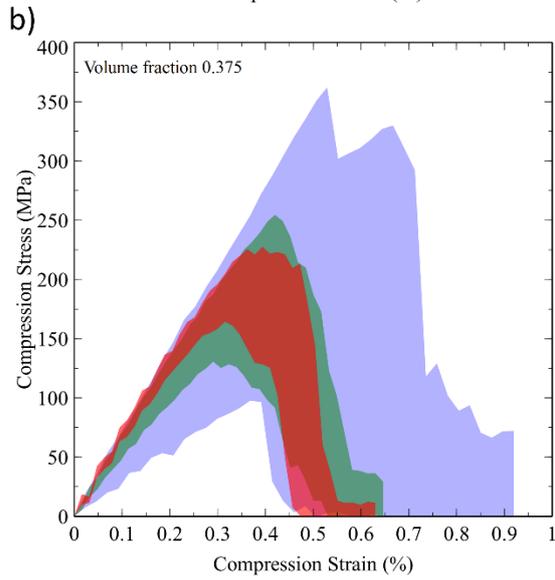
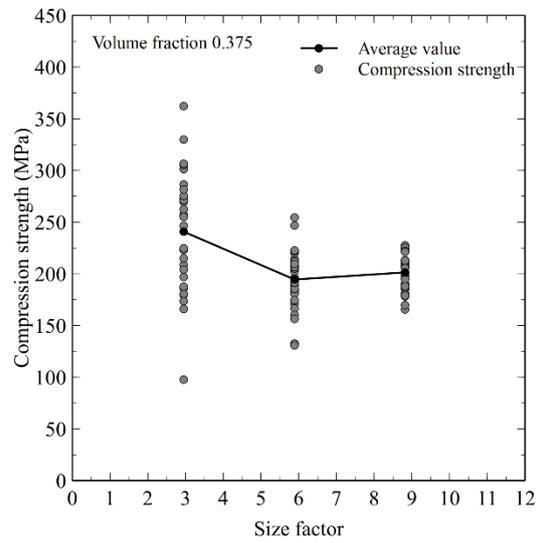

c)
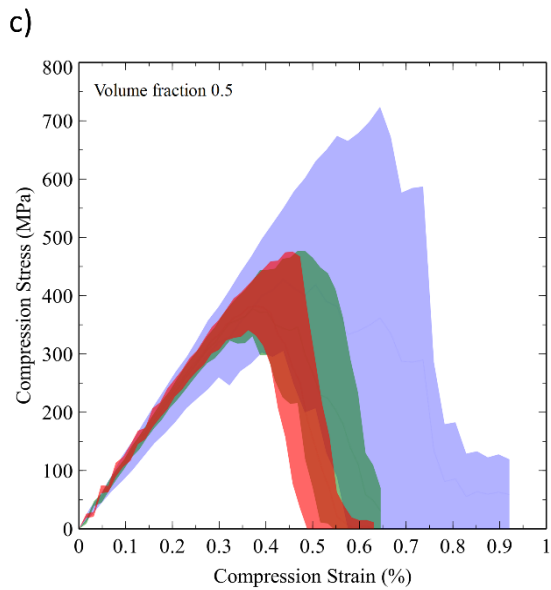
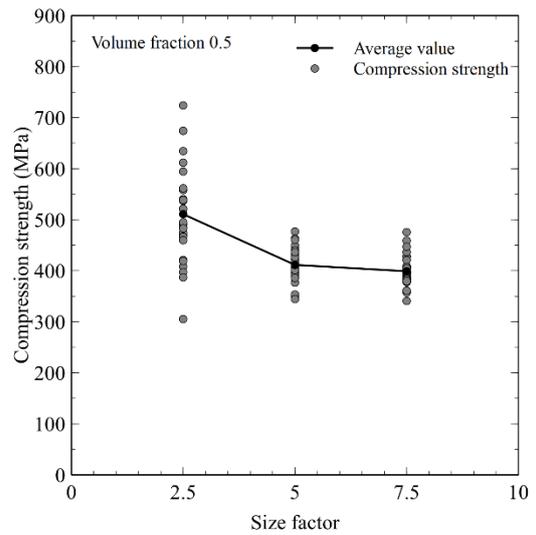

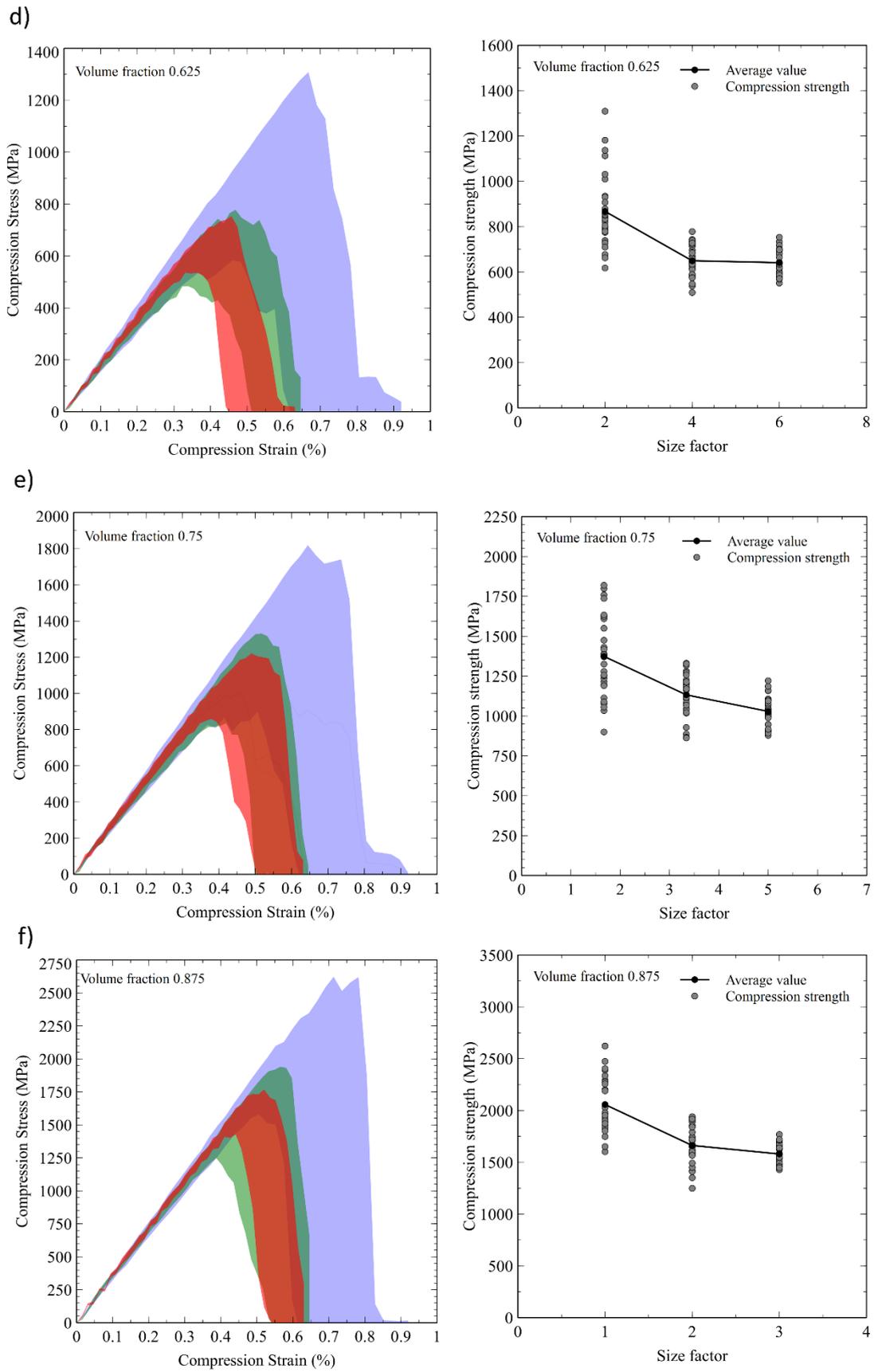

Fig. 13: The left-side image shows scatter in compression stress-strain behaviour and the right-side image shows scatter in compression strength value of VEs for three different sizes and for volume fraction a) 0.125, b) 0.375, c) 0.5, d) 0.625, e) 0.75 and f) 0.875.

| Alumina volume fraction of VE | Edge length 50 (voxels) | | Edge length (100 voxels) | | Edge length (150 voxels) | |
|---|---|---|---|---|---|---|
| | Compression strength (MPa) | Coefficient of variation (%) | Compression strength (MPa) | Coefficient of variation (%) | Compression strength (MPa) | Coefficient of variation (%) |
| 0.125 | 7.56 | 54.32 | 6.11 | 18.04 | 7.2 | 11.5 |
| 0.255 | 76.51 | 33.25 | 71.66 | 13.98 | 70.7 | 9.74 |
| 0.375 | 240.78 | 23.26 | 194.52 | 14.30 | 201.36 | 8.50 |
| 0.5 | 510.84 | 17.81 | 411.49 | 8.35 | 398.97 | 7.31 |
| 0.625 | 867.16 | 18.11 | 649.24 | 10.65 | 640.81 | 8.67 |
| 0.75 | 1373.67 | 17.75 | 1132.29 | 11.40 | 1028.65 | 8.69 |
| 0.875 | 2057.17 | 13.04 | 1662.15 | 10.66 | 1579.78 | 5.73 |

Table 3: Average compression strength and coefficient of variation for VEs of each size and for every volume fraction of alumina.

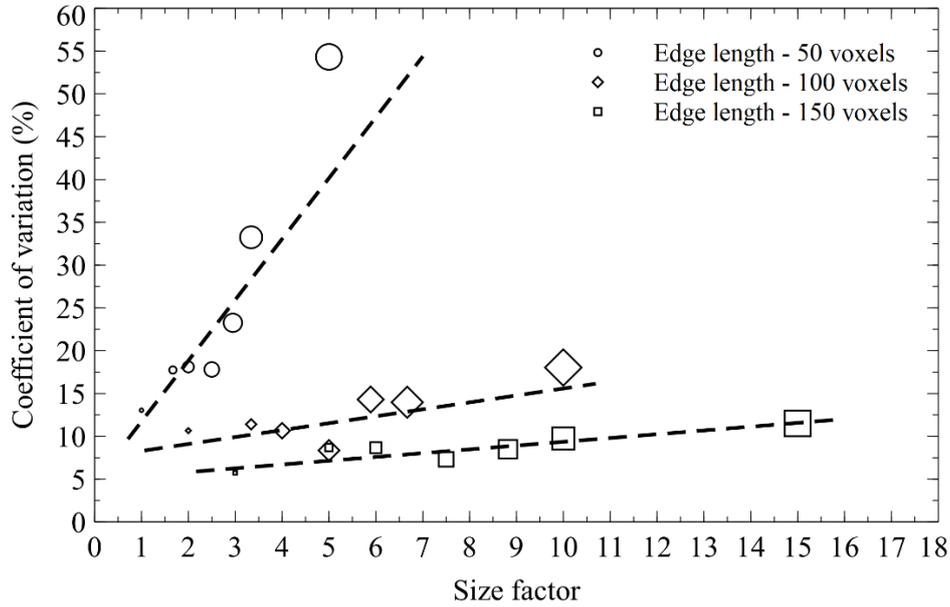

Fig. 14: Coefficient of variation with respect to size factor for three sizes of VEs. Note that in each VE size, the smallest marker denotes result of largest volume fraction 0.875 and the largest one denotes that of smallest volume fraction 0.125.

Table. 3 shows the values of average compression strength and coefficient of variation for VEs of each size and volume fraction of alumina. Fig.14 depicts how the coefficient of variation changes with size factor for each size of VE and volume fraction. Note that the smallest marker in each VE size indicate the result of the largest volume fraction and the largest marker indicate the result of the smallest volume fraction. The shadow curves of compression stress-strain for VEs with edge length 150 voxels is shown in Fig. 15a for all selected volume fractions. For the same VEs, Fig. 15b shows variation of average compression strength with solid volume fraction. The results are compared with analytical estimates for cellular solids described in [10], [11] also known as the Gibson-Ashby model. These are proportionality laws that define relationships between effective material property of cellular solids and the relative density. The compression strength relation developed for open-cell and closed-cell cellular solids is given in Eq. (6)

$$\sigma_s = C \sigma_{base} \left(\frac{\rho_{cellular}}{\rho_{base}}\right)^n \qquad (6)$$

Here, $\sigma_s$ and $\sigma_{base}$ indicate compression strength of cellular solid and base material respectively. $\rho_{foam}$ and $\rho_{base}$ indicate density of cellular solid and base material respectively. The exponent $n$ takes value of $\frac{3}{2}$ for open-cell and value of 2 for closed-cell cellular solids. The constant $C$ was predicted as 0.65 from experimental measurements in [10].

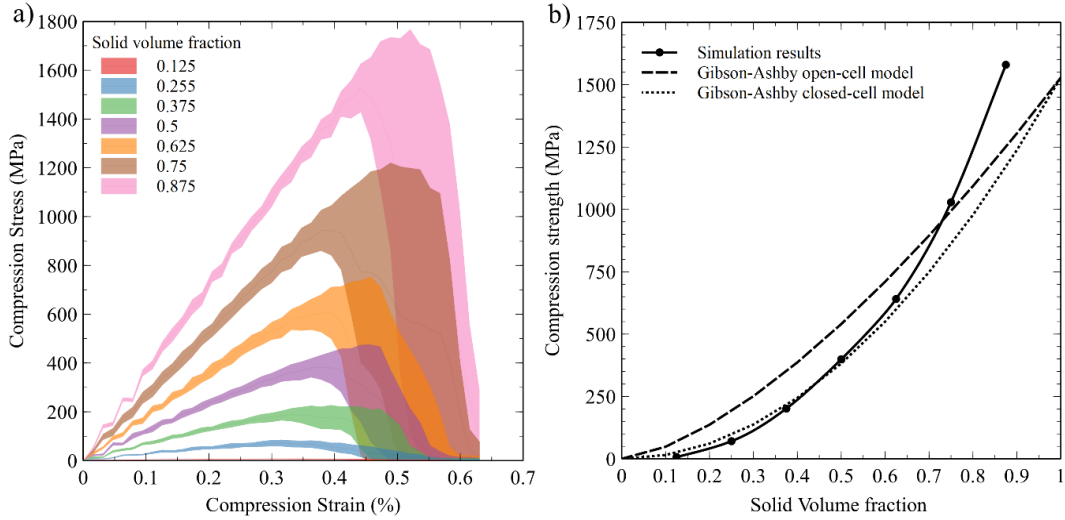

Fig. 15: a) Shadow curves of compression stress-strain for VEs of edge length 150 voxels and all volume fractions; b) variation of average compression strength with volume fraction along with results of Gibson-Ashby models.

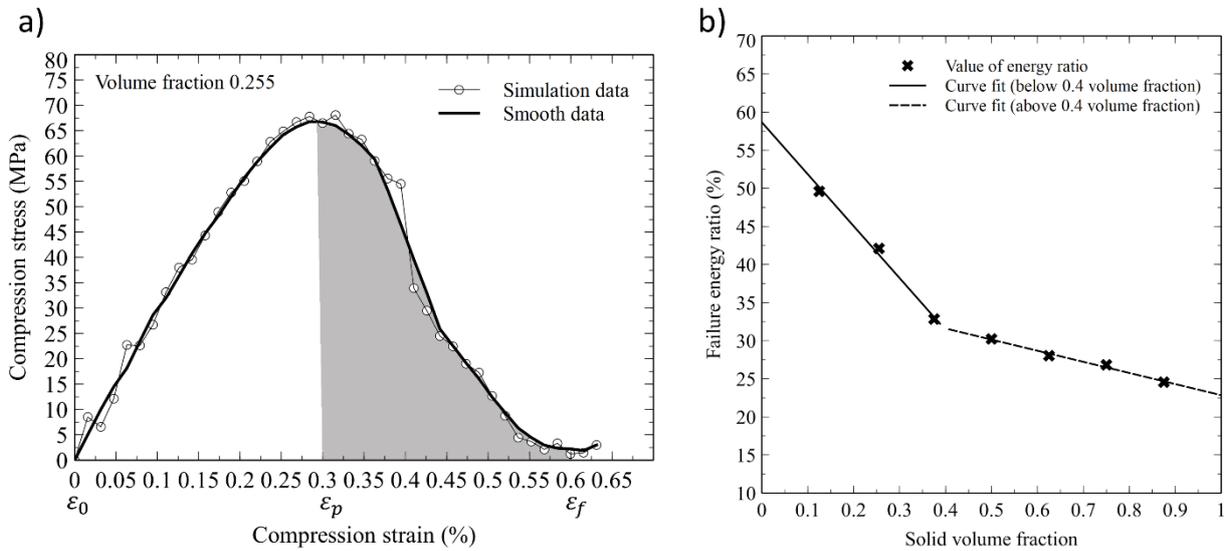

Fig. 16: a) Simulation and smoothed compression stress-strain curve of a VE of edge length 150 voxels and volume fraction 0.255; b) variation of failure energy ratio with volume fraction.

In order to study the energy absorption behaviour of the microstructure during compression, the area under the stress-strain curve is calculated. Fig. 16a shows compression stress-strain curve (simulation and smoothed data) of a VE of edge length 150 voxels for microstructure with volume fraction 0.255. The complete area under the stress-strain curve represents the total energy absorbed by the VE during the compression. It is denoted by $E_T$. The shaded area in Fig. 16a represents the energy absorbed post peak stress. Let us name it as failure energy denoted by $E_f$. The expressions for these energies are given in Eq. (7) and (8). $\varepsilon_0$, $\varepsilon_p$ and $\varepsilon_f$ denote initial strain, strain at peak stress and strain at complete failure respectively. The ratio of failure energy to total energy indicates the portion of the total energy absorbed post peak stress. This failure energy ratio denoted by $R_{fe}$ is expressed in Eq. (9).

$$E_T = \int_{\varepsilon_0}^{\varepsilon_f} \sigma(\varepsilon)\, d\varepsilon \tag{7}$$

$$E_f = \int_{\varepsilon_p}^{\varepsilon_f} \sigma(\varepsilon)\, d\varepsilon \tag{8}$$

$$R_{fe} = \frac{E_f}{E_T} \times 100 \tag{9}$$

This ratio is calculated for all the VEs of the highest size and for each volume fraction. The ensemble average is then plotted with respect to the volume fraction as shown in Fig. 16b. It can be seen that as the volume fraction is increased, the failure energy ratio is decreased. However, there is a distinct change in the nature of this reduction as the volume fraction is increased from 0.375 to 0.5. The rate of reduction is drastic as the volume fraction is increased from 0.125 to 0.375. However, beyond that the reduction is not as significant.

### 3.4 Nature of damage propagation in microstructures with different volume fractions

When subjected to compressive loading, the nature of damage propagation in the microstructure depends on the volume fraction of the solid content present in it. In order to study this, the evolution of damage in the VEs of each volume fraction is studied. For the case of volume fraction 0.255, this study has already been described in section 2.1. In the present section, VE of volume fraction 0.875 is studied as it shows a different nature of damage propagation as compared to the earlier case.

Fig. 17a shows the compression stress-strain curve of a VE with edge length 150 voxels and volume fraction 0.875. The direction of loading and the final damage pattern at the end of simulation in shown in Fig. 17b. In order to study the nature of damage propagation, four intermediate stages of the simulation course, marked as A to D are shown in Fig. 17a. As the compressive strain is increased, the microstructure undergoes elastic deformation. The formation of first damaged regions is observed at stage A at the strain value of 0.19%. The two locations where damage is initiated has been marked in dotted circles in Fig. 17c. These locations are enlarged and shown in the adjacent images. At location $A$, the damage (red color region) is initiated at the top and/or the bottom edge of the three pores. At location $B$, the damage is initiated in a very thin strut that is formed in the region where the three neighbouring pores come very close to each other. As the strain is increased, damage is formed in the direction of the loading along the circumference of the biggest pores present in the microstructure. This can be seen in Fig.17d where the red regions seem to trace circles. This is stage B at the strain value of 0.4%. The maximum compressive stress is reached at stage C at the strain value of 0.46%. The damage pattern at this stage can be seen in Fig. 17e. At this stage the damage that had initiated along the circumference of the neighbouring pores has merged to form a macro crack or a macro damaged zone. In Fig. 17e, the connected red color circular regions can be seen in the vertical plane as well as in the horizontal plane (the right-side image). At stage D at the strain value of 0.58%, the sample shows complete loss of strength. Fig. 17f shows the damage pattern at this stage which indicates that the damage has grown to the extent that there is an abrupt failure in a plane that had weakened due to the existence of the previous damaged regions. The same damage pattern along with undamaged foam region can be seen in Fig. 17b. It can be seen that initially, the damage propagated vertically from the pores till the adjacent damaged regions interacted with each other to form a macro crack that resulted in an abrupt failure. The sample at this stage is broken abruptly into two smaller pieces.

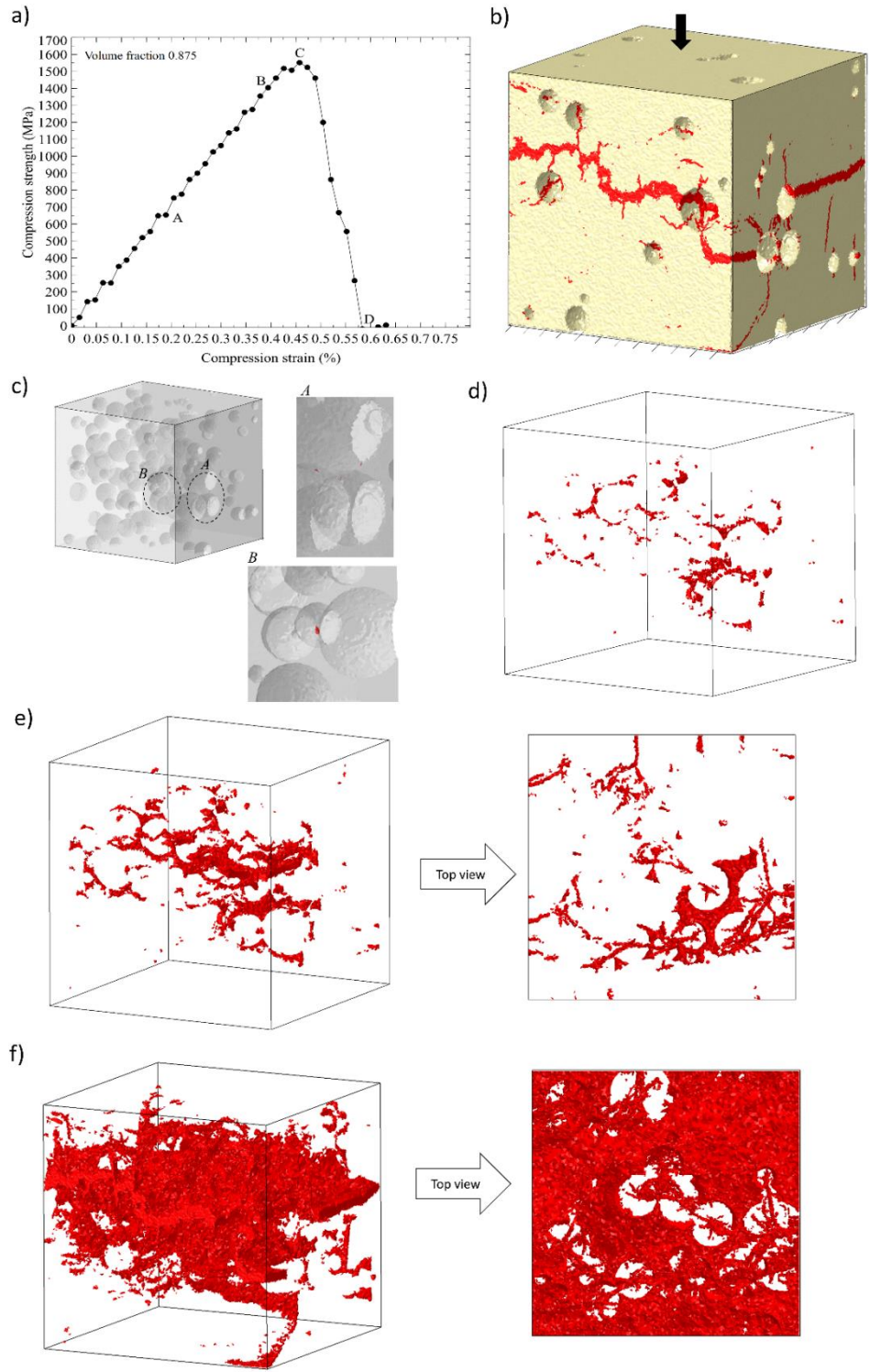

Fig. 17: a) Effective compression stress-strain curve of VE with edge length 150 voxels and volume fraction 0.875; b) final damage pattern in VE; c) location of initial damage in the VE; damage pattern at d) stage B, e) stage C, f) stage D of the simulation course.

A similar study is conducted on VEs of all the remaining volume fractions. The left side images in Figs. 18a-g show final damage pattern in VEs of volume fraction 0.125,0.255,0.375,0.5,0.625,0.75 and 0.875 respectively. The right-side images show the broken parts of the VE when it has lost all its strength. The yellow-coloured regions are the parts of the VE that have been completely detached from the main body (green coloured-region). Note that there are many small parts that have completely lost their load bearing ability but are still attached to the main body. The yellow-coloured regions do not represent such parts. The objective behind such illustration is to show that VEs with volume fraction 0.125 to 0.375 (Fig. 18a-c) fail in a cellular fashion. It means that the VEs get disintegrated into several small parts as the struts fail. Beyond volume fraction of 0.375, the VEs show abrupt brittle failure. The right-side images in Figs. 18d-g show that the VE breaks into two major parts shown in yellow and green color.

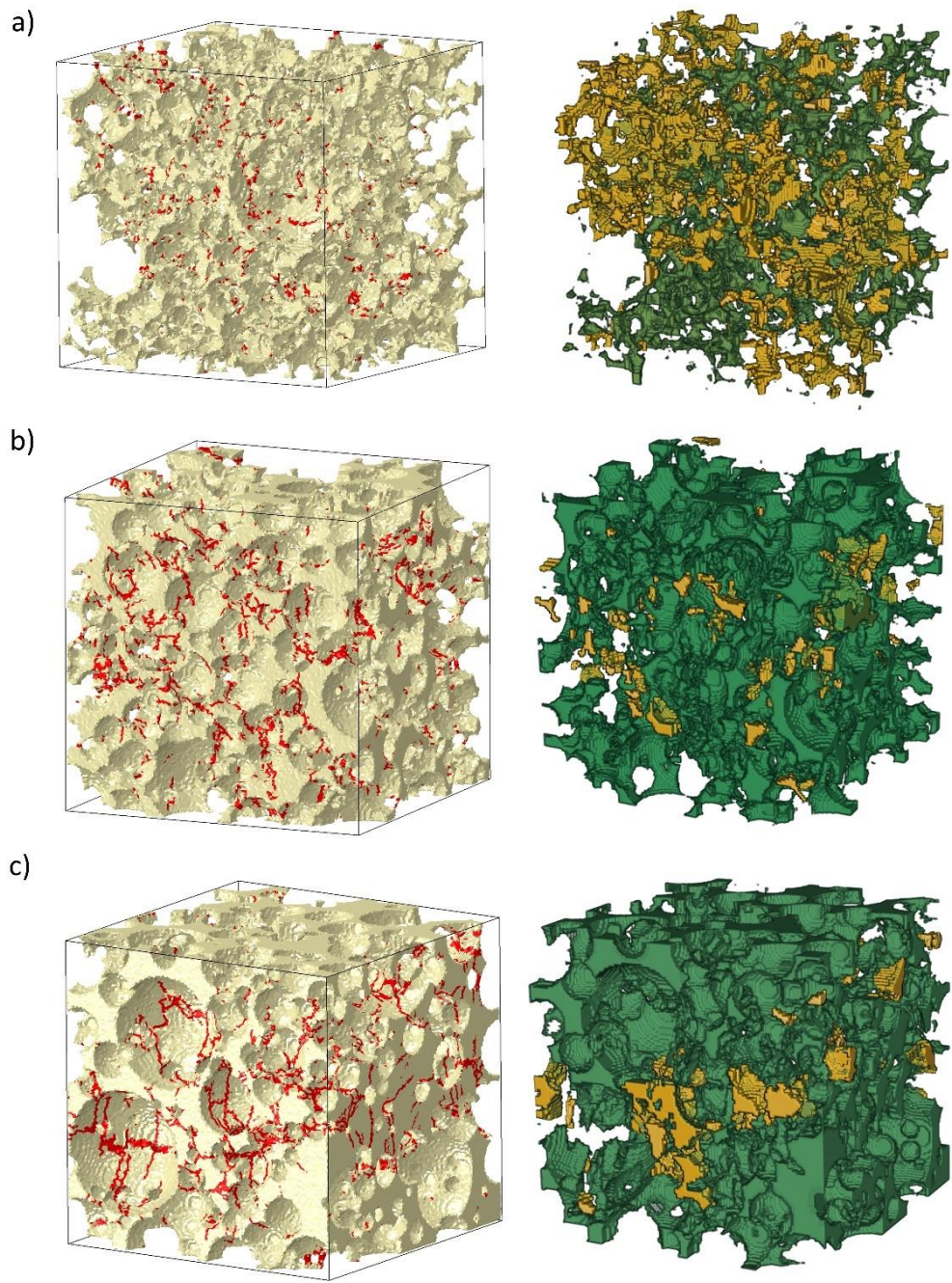

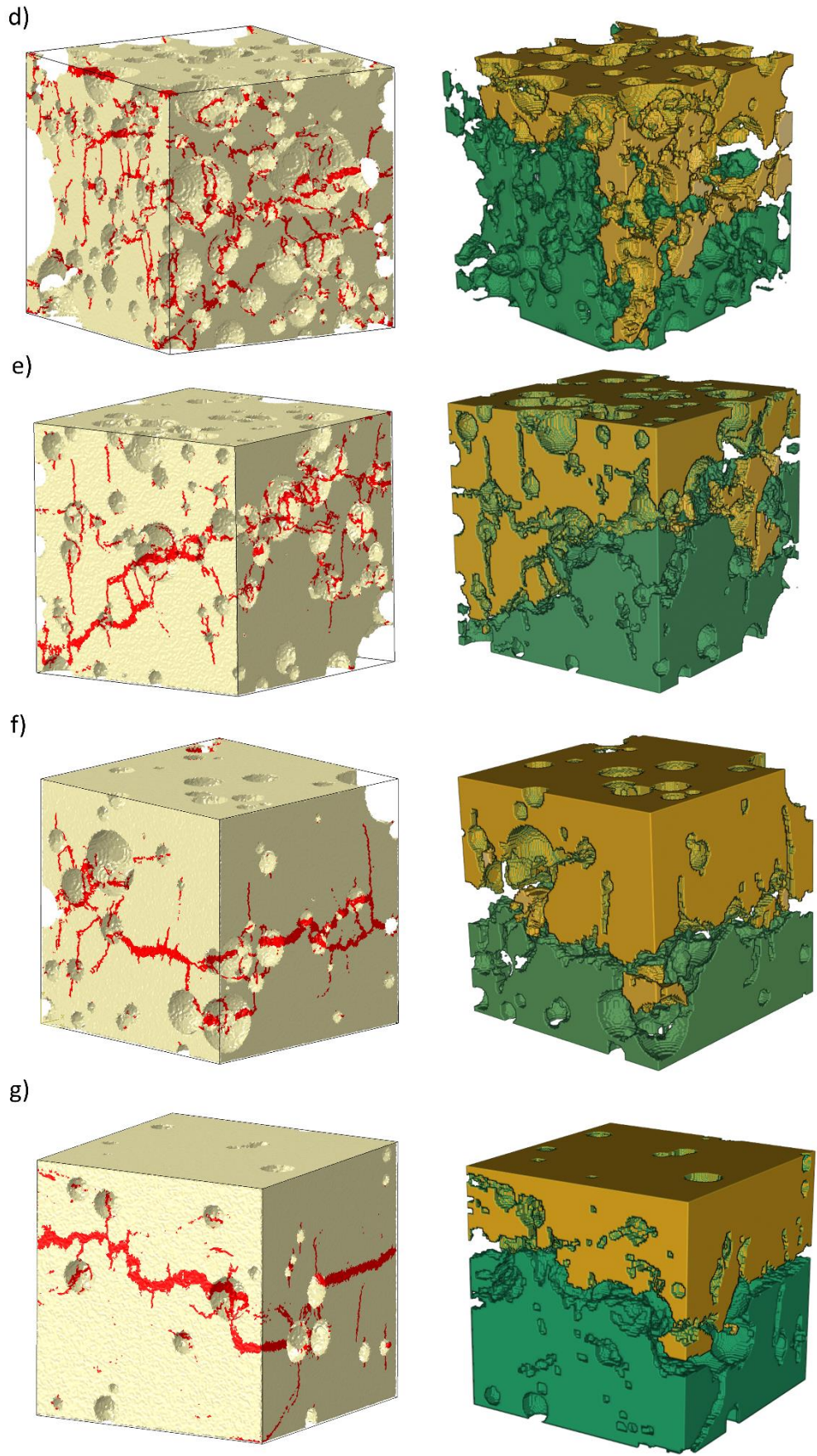

Fig. 18: The left-side image shows pattern of damage propagation and the right-side image shows breakage when the VE losses complete strength for volume fraction a) 0.125, b) 0.255, c) 0.375, d) 0.5, e) 0.625 and f) 0.75 and g) 0.875

## 4. Discussion

In this work, an attempt is made to answer the research question of how does a compressive failure behaviour of porous ceramics varies with its solid content. In our earlier published article [17], a microstructure reconstruction algorithm was proposed that could develop artificial microstructures of porous materials when the shape and size distribution of pores, target statistical correlation functions and the volume fraction were given. This algorithm forms the basis for creation of microstructures with different volume fractions in the present work.

As a reference material system, a real foam sample with volume fraction 0.255 is chosen. The microstructure in Fig. 1a can be seen as a random distribution of spherical pores penetrating each other. A representative volume element of the sample identified in [17] is chosen to perform finite element simulation of the compressive failure. Fig. 3 shows that the simulation results in three orthogonal directions are in agreement with the experimental results. It can be seen in Fig. 2 that the microstructure is formed by a network of thin struts formed around the pores. It is observed that the microstructure fails by bending of struts. Since the struts are distributed throughout the sample and do not have any particular orientation, the damage is also distributed and does not form any localized failure zone. This is a typical behaviour of cellular materials as described in the works [10][24] where proportionality laws known as Gibson-Ashby models were developed to relate effective foam properties with the relative density. They utilized a simplified model of foam as shown in Fig.19a. A cubic unit cell was assumed with the solid material distributed as a network of beams in open-cell model and plates in a closed-cell model. The hypothesis was that the macroscopic compressive failure of foam is directly linked to the failure of beams (in open-cell) and plates (in closed-cell) in bending. This assumption bodes well with the observations from the simulation of real foam sample in the present work.

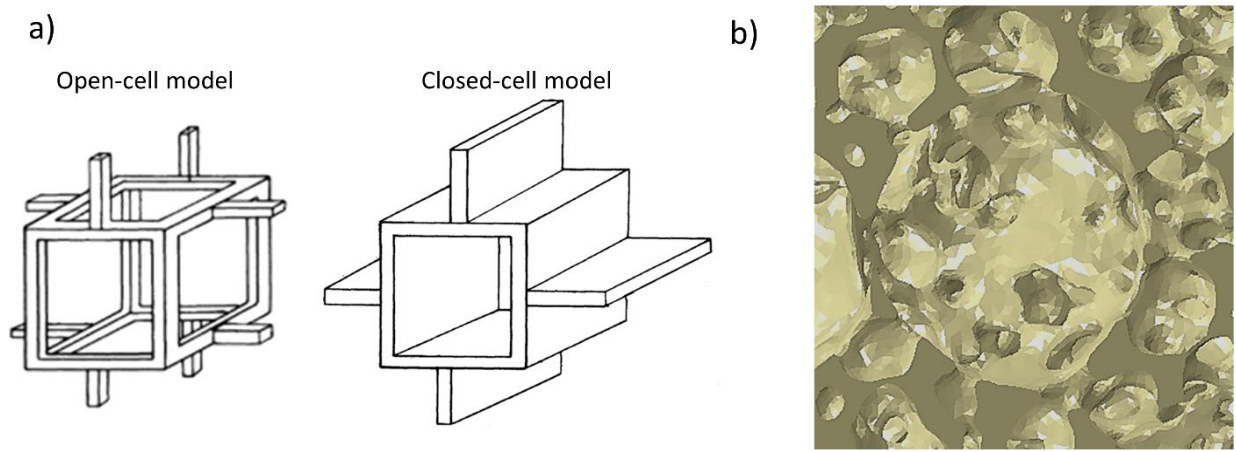

Fig. 19: a) Open-cell and closed-cell model as defined in [10] [24]; b) microstructure of real foam sample focussing on the surface of a typical pore

One important reason behind recreating microstructures is to study the effect of VE size on the compression failure behaviour. Previous studies in this regard were focused on cellular materials whose macroscopic dimensions were comparable to their characteristic lengths. [12], [13] developed analytical models based on 2D honeycombs with the assumption that the peripheral cells, near the free edges, are not loaded as much as the inner cells. They developed relations that state that the compression strength decreases with decrease in specimen size and proved this with experimental studies. The follow up article [14] concluded that this size effect is mainly because of the structural defects present in the foams and would not exists otherwise. In the present study, although the size difference between the microscopic and macroscopic dimensions would be of the order of 3 or more, it is important to determine appropriate size of a VE that can be used to calculate the effective material properties. In section 2.2, artificial microstructures equivalent to the real foam sample were generated for 3 different sizes as shown in Fig. 7a. The shadow diagrams for simulated stress-strain curves (refer Fig. 8a) along with the scatter in the compression strength values (refer Fig. 8b) show that the average compression behaviour does not change much with VE size but the scatter in the results decreases drastically as the size in increased. These results agree with the observations in [14].

Now that the numerical procedure has been established, the next step is to generate artificial microstructures for different volume fractions. Section 3.1 describes the method adopted to provide all the inputs required for the reconstruction algorithm. The reconstructed VEs shown in Fig. 12 are utilized in finite element simulations to study compression stress-strain behaviour. The left-side images in Fig. 13 shows the shadow diagrams for each volume fraction. The nature of the shadow diagrams is similar to that shown in Fig. 8a. The scatter is very large for the case of the VE size of 50 voxels but it reduces drastically for 100 voxels. The reduction in scatter is not as much from 100 to 150 voxels. The average compression strength values and their scatter are shown in the right-side images of Fig. 13. For all the studied volume fractions, the change in the average strength values is not significant when the VE size is increased but the reduction in the scatter is noticeable. The values of the average compression strength and the coefficient of variation are shown in Table.3. It is observed in Table.2 that as the volume fraction is increased, the size factor is decreased for all VE sizes. More understanding can be derived from Fig. 14 which shows that for the VE size of 50 voxels, the coefficient of variation is highly sensitive to the size factor and hence the volume fraction. This sensitivity reduces drastically for the VE size of 100 voxels and is almost independent of the size factor (and volume fraction) for the VE size of 150 voxels. Another important inference is that for the volume fraction of 0.875, the coefficient of variation is least affected by the VE size. But as the volume fraction is reduced, its value becomes more and more dependent on the VE size. With these observations, it was decided that for studying the effect of volume fraction on the compression strength, the VEs of size 150 voxels would be utilized.

Figs.15a and b show variation of compression failure behaviour with volume fraction for the above selected VE size. In Fig. 15b, the comparison of simulation results with that of the analytical results obtained from the Gibson-Ashby models bring forth two important observations. First, the simulation results appear to agree well with the closed-cell model than the open-cell model. This can be justified from Fig. 19b which shows pore surface of a typical pore in a foam microstructure. The surface resembles a closed surface with holes in between rather than a network of connected beams. Although the Gibson-Ashby model was formulated with a cubical pore (refer Fig.19a), it provides good estimates to real foam microstructures with rounded pores as shown in their work in [24]. The second observation is that the simulation results start diverging from the closed-cell analytical results as the volume fraction increases beyond 0.5. The understanding behind this is that there is a change in the mode of failure as the volume fraction increases beyond a certain value. When the volume fraction is less, the material behaves as a cellular solid with the damage propagation as described in Fig.4 for the real foam sample. As the volume fraction is increased beyond a certain value, the material behaves as a brittle solid with a few pores distributed within it.

An example of this second mode of failure is shown in Fig. 17 for the microstructure with volume fraction 0.875. Contradictory to cellular solids, the damage in this case initiates at specific sites of stress concentration. The damage also initiates at the top and bottom regions along the circumference of the biggest holes as these regions are subjected to maximum tensile stress. These are the typical sites of damage initiation when a plate with a hole is loaded in compression as described in the theoretical works of [8], [9]. As the loading is increased, the damaged zones along the circumference of the neighbouring pores merge to form a macro damaged zone. This is followed by an abrupt shear failure when the damage increases rapidly along the energetically most preferred plane.

Coming back to Fig. 15b, the Gibson-Ashby models were formulated to work only for small volume fractions where the struts fail by bending and not by compression or shear which occur when the volume fraction is large. They predicted the transition value of volume fraction to be around 0.6 (refer [24]). These two failure modes can be observed in Fig.16b which shows clear transition around the volume fraction of 0.375 - 0.5. Below this value, the percentage of energy absorbed after peak stress is high and it increases drastically as the volume fraction is deceased (or as the porosity is increased). This suggests that the material failure is gradual as observed in cellular materials. Above the volume fraction of 0.5, the percentage of energy absorbed after the peak stress is low and is not as much dependent on the volume fraction either indicating that the material failure is abrupt and irrespective of the volume fraction. The damage pattern and the damaged state of the VEs shown in the left-side and the right-side images of Fig. 18 respectively provide visual validation to these observations. The left-side images of figs. 18a-c, for volume fraction 0.125-0.375, show that the damage pattern is distributed throughout the VE. The final damaged state of the VEs (right-side images) show small yellow-coloured parts totally detached from the main body. There are also many more small parts that have lost their strength completely but are still attached to the main body. This is the results of struts failure homogeneously distributed across the sample. Fig. 18d for volume fraction 0.5, shows transitional mode of failure. The damage pattern shows struts failure throughout the space but also shows formation of a macro-damaged zone formed by merging of small damaged areas. The right-side image shows failure of the VE into two major parts, indicative of a localized failure but also small parts broken away from the main body. Figs. 19e-g for volume fraction 0.625 – 0.875 show abrupt failure by shear. The left-side images show vertical damage lines formed due to tensile stress on the top and bottom regions of the pore circumference. The final failure occurs by the formation of a macro-damaged zone that splits the VE into two. The right-side images show that the VEs have been split into two major parts separated by the macro-damaged zone. Figs. 20a and 20b show illustrations of the cellular and brittle damage respectively. The black dotted lines indicate failed sturts. In the case of cellular failure (Fig.20a), the black dotted lines are distributed homogeneously throughout the volume. On the contrary, in the case of brittle failure (Fig.20b), they are mostly concentrated in a localized zone that leads to formation of a preferred plane of failure by shear as shown by a light grey-coloured plane. These findings are in agreement with the experimental observations reported in [2] which studied compressive failure of alumina foams for a range of volume fractions. They observed that the samples with volume fraction less than 0.5 failed in a cellular fashion while those with volume fraction more than 0.5 showed brittle failure. In case of brittle failure, long cracks were observed followed by crack-crack interaction to form macrocracks that cause final failure. In the case of cellular failure, the cell walls (struts) between the pores failed throughout the sample.

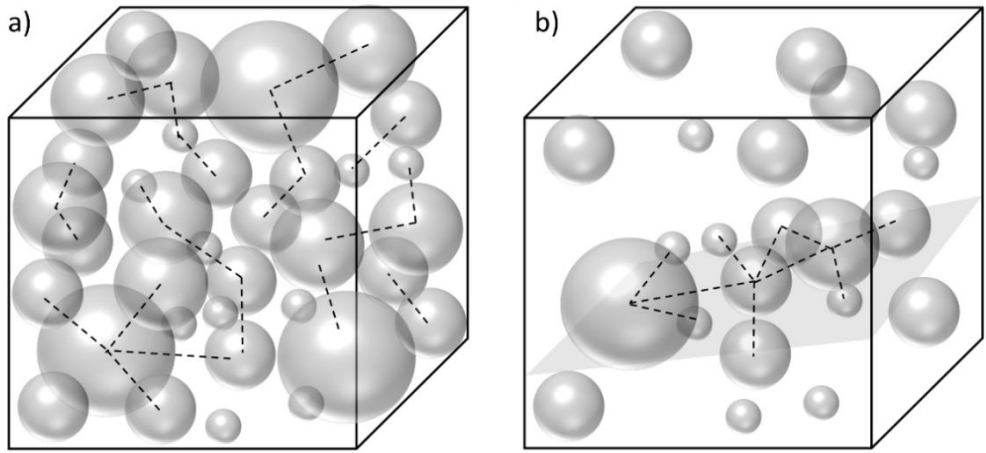

Fig. 20: Illustration of a) cellular failure and b) brittle failure in porous ceramics

## 5. Conclusion

The work focuses on studying the compression failure behaviour of cellular ceramics through reconstruction of artificial microstructures and finite element simulations. The simulation result of one volume fraction is compared with that of the experimental measurements to establish fidelity of the numerical procedure. For each volume fraction, effect of volume element size on compression failure behaviour is studied and a representative size is identified. By simulating multiple samples for each size and volume faction, the scatter in compression stress-stress curves and compression strength is studied. The simulation results correlated well with the analytical estimates of the Gibson-Ashby model and also reinforce the assumptions made in the model. The two modes of failure observed in the simulations namely: cellular failure below 0.5 volume fraction and brittle abrupt failure above it concurred with the experimental observations in [2]. This study forms the basis of the future efforts in which we plan to develop multiscale simulation procedures to study the compression failure behaviour of porous ceramics at a macroscale level.

[1] A. R. Studart, U. T. Gonzenbach, E. Tervoort, and L. J. Gauckler, "Processing Routes to Macroporous Ceramics: A Review," *J. Am. Ceram. Soc.*, vol. 89, no. 6, pp. 1771–1789, Jun. 2006, doi: 10.1111/J.1551-2916.2006.01044.X.

[2] S. Meille, M. Lombardi, J. Chevalier, and L. Montanaro, "Mechanical properties of porous ceramics in compression: On the transition between elastic, brittle, and cellular behavior," *J. Eur. Ceram. Soc.*, vol. 32, no. 15, pp. 3959–3967, Nov. 2012, doi: 10.1016/J.JEURCERAMSOC.2012.05.006.

[3] P. Colombo and M. Modesti, "Silicon Oxycarbide Ceramic Foams from a Preceramic Polymer," *J. Am. Ceram. Soc.*, vol. 82, no. 3, pp. 573–578, Mar. 1999, doi: 10.1111/J.1151-2916.1999.TB01803.X.

[4] P. Colombo, J. R. Hellmann, and D. L. Shelleman, "Mechanical Properties of Silicon Oxycarbide Ceramic Foams," *J. Am. Ceram. Soc.*, vol. 84, no. 10, pp. 2245–2251, Oct. 2001, doi: 10.1111/J.1151-2916.2001.TB00996.X.

[5] B. S. M. Seeber, U. T. Gonzenbach, and L. J. Gauckler, "Mechanical properties of highly porous alumina foams," *J. Mater. Res.*, vol. 28, no. 17, pp. 2281–2287, Sep. 2013, doi: 10.1557/JMR.2013.102.

[6] C. Voigt, J. Storm, M. Abendroth, C. G. Aneziris, M. Kuna, and J. Hubálková, "The influence of the measurement parameters on the crushing strength of reticulated ceramic foams," *J. Mater. Res.*, vol. 28, no. 17, pp. 2288–2299, Sep. 2013, doi: 10.1557/JMR.2013.96.

[7] R. J. Mora and A. M. Waas, "Strength scaling of brittle graphitic foam," *Proc. R. Soc. London. Ser. A Math. Phys. Eng. Sci.*, vol. 458, no. 2023, pp. 1695–1718, Jul. 2002, doi: 10.1098/RSPA.2001.0938.

[8] C. G. Sammis and M. F. Ashby, "The failure of brittle porous solids under compressive stress states," *Acta Metall.*, vol. 34, no. 3, pp. 511–526, Mar. 1986, doi: 10.1016/0001-6160(86)90087-8.

[9] M. F. Ashby and C. G. Sammis, "The damage mechanics of brittle solids in compression," *pure Appl. Geophys. 1990 1333*, vol. 133, no. 3, pp. 489–521, May 1990, doi: 10.1007/BF00878002.

[10] S. K. Maiti, L. J. Gibson, and M. F. Ashby, "Deformation and energy absorption diagrams for cellular solids," *Acta Metall.*, vol. 32, no. 11, pp. 1963–1975, Nov. 1984, doi: 10.1016/0001-6160(84)90177-9.

[11] L. J. Gibson and M. F. Ashby, "Cellular solids : structure and properties," *Cambridge Univ. Press*, p. 510, 1997.

[12] P. R. Onck, E. W. Andrews, and L. J. Gibson, "Size effects in ductile cellular solids. Part I: modeling," *Int. J. Mech. Sci.*, vol. 43, no. 3, pp. 681–699, Dec. 2001, doi: 10.1016/S0020-7403(00)00042-4.

[13] E. W. Andrews, G. Gioux, P. Onck, and L. J. Gibson, "Size effects in ductile cellular solids. Part II: experimental results," *Int. J. Mech. Sci.*, vol. 43, no. 3, pp. 701–713, Dec. 2001, doi: 10.1016/S0020-7403(00)00043-6.

[14] C. Tekoğlu, L. J. Gibson, T. Pardoen, and P. R. Onck, "Size effects in foams: Experiments and modeling," *Prog. Mater. Sci.*, vol. 56, no. 2, pp. 109–138, Feb. 2011, doi: 10.1016/J.PMATSCI.2010.06.001.

[15] D. Horny, J. Schukraft, K. A. Weidenmann, and K. Schulz, "Numerical and Experimental Characterization of Elastic Properties of a Novel, Highly Homogeneous Interpenetrating Metal Ceramic Composite," *Adv. Eng. Mater.*, vol. 22, no. 7, p. 1901556, Jul. 2020, doi: 10.1002/ADEM.201901556.

[16] J. Schukraft, D. Horny, K. Schulz, and K. A. Weidenmann, "3D modelling and experimental investigation on the damage behavior of an interpenetrating metal ceramic composite (IMCC) under compression," *Mater. Sci. Eng. A*, p. 143147, Apr. 2022, doi: 10.1016/J.MSEA.2022.143147.

[17] V. V. Deshpande, K. A. Weidenmann, and R. Piat, "Application of statistical functions to the numerical modelling of ceramic foam: From characterisation of CT-data via generation of the virtual microstructure to estimation of effective elastic properties," *J. Eur. Ceram. Soc.*, vol. 41, no. 11, pp. 5578–5592, 2021, doi: 10.1016/j.jeurceramsoc.2021.03.054.

[18] "MATLAB - MathWorks - MATLAB & Simulink." https://www.mathworks.com/products/matlab.html?s_tid=hp_ff_p_matlab (accessed May 02, 2022).

[19] "Abaqus Unified FEA - SIMULIA$^{TM}$ by Dassault Systèmes®." https://www.3ds.com/products-services/simulia/products/abaqus/ (accessed Apr. 24, 2022).

[20] G. R. Johnson and T. J. Holmquist, "An improved computational constitutive model for brittle materials," in *AIP Conference Proceedings*, May 1994, vol. 309, no. 1, pp. 981–984. doi: 10.1063/1.46199.

[21] G. Guo, S. Alam, and L. D. Peel, "Numerical analysis of ballistic impact performance of two ceramic-based armor structures," *Compos. Part C Open Access*, vol. 3, p. 100061, Nov. 2020, doi: 10.1016/J.JCOMC.2020.100061.

[22] W. K. Schukraft J, Lohr C, "Mechanical characterization of an interpenetrating metal-matrix composite based on highly homogeneous ceramic foams," *Hybrid 2020- Mater. Struct. Proc.*, pp. 33–39, 2020.

[23] Y. Jiao, F. H. Stillinger, and S. Torquato, "Modeling heterogeneous materials via two-point correlation functions. II. Algorithmic details and applications," *Phys. Rev. E - Stat. Nonlinear, Soft Matter Phys.*, vol. 77, no. 3, p. 031135, Mar. 2008, doi: 10.1103/PHYSREVE.77.031135/FIGURES/19/MEDIUM.

[24] L. J. Gibson and M. F. Ashby, "The mechanics of three-dimensional cellular materials," *Proc. R. Soc. London. A. Math. Phys. Sci.*, vol. 382, no. 1782, pp. 43–59, Jul. 1982, doi: 10.1098/RSPA.1982.0088.